# Probing magnetic ordering in air stable iron-rich van der Waals minerals

*Muhammad Zubair Khan[1,\*], Oleg E. Peil[2], Apoorva Sharma[3], Oleksandr Selyshchev[3,6], Sergio Valencia[4], Florian Kronast[4], Maik Zimmermann[5], Muhammad Awais Aslam[1], Johann G. Raith[5], Christian Teichert[1], Dietrich R.T. Zahn[3,6], Georgeta Salvan[3,6], and Aleksandar Matković[1,\*]*

[1] Chair of Physics, Department Physics, Mechanics and Electrical engineering, Montanuniversität Leoben, 8700, Leoben, Austria.
[2] Materials Center Leoben Forschung GmbH (MCL), 8700, Leoben, Austria.
[3] Semiconductor Physics, Chemnitz University of Technology, D-09107, Chemnitz, Germany.
[4] Department of Spin and Topology in Quantum Materials, Helmholtz-Zentrum Berlin, Albert-Einstein-Str. 15, D-12489, Berlin, Germany.
[5] Chair of Resource Mineralogy, Montanuniversität Leoben, 8700, Leoben, Austria.
[6] Centre for Materials, Architecture, and Integration of Nanomembranes (MAIN), Chemnitz University of Technology, 09126, Chemnitz, Germany.
Corresponding email: muhammad.khan@unileoben.ac.at, aleksandar.matkovic@unileoben.ac.at

**Abstract**

In the rapidly expanding field of two-dimensional materials, magnetic monolayers show great promise for the future applications in nanoelectronics, data storage, and sensing. The research in intrinsically magnetic two-dimensional materials mainly focuses on synthetic iodide and telluride based compounds, which inherently suffer from the lack of ambient stability. So far, naturally occurring layered magnetic materials have been vastly overlooked. These minerals offer a unique opportunity to explore air-stable complex layered systems with high concentration of local moment bearing ions.

We demonstrate magnetic ordering in iron-rich two-dimensional phyllosilicates, focusing on mineral species of minnesotaite, annite, and biotite. These are naturally occurring van der Waals magnetic materials which integrate local moment baring ions of iron via magnesium/aluminium substitution in their octahedral sites. Due to self-inherent capping by silicate/aluminate tetrahedral groups, ultra-thin layers are air-stable. Chemical characterization, quantitative elemental analysis, and iron oxidation states were determined via Raman spectroscopy, wavelength disperse X-ray spectroscopy, X-ray absorption spectroscopy, and X-ray photoelectron spectroscopy. Superconducting quantum interference device magnetometry measurements were performed to examine the magnetic ordering. These layered materials exhibit paramagnetic or superparamagnetic characteristics at room temperature. At low temperature ferrimagnetic or antiferromagnetic ordering occurs, with the critical ordering temperature of 38.7 K for minnesotaite, 36.1 K for annite, and 4.9 K for biotite. In-field magnetic force microscopy on iron bearing phyllosilicates confirmed the paramagnetic response at room temperature, present down to monolayers. Our results unveil a new class of magnetic insulators with ambient stability, which enable up to 100 percent substitution of the core ions with the magnetic moment baring species.

## 1 Introduction:

The discovery of graphene[1] sparked the development of many other two-dimensional (2D) van der Waals (vdW) materials.[2] Driven by the needs in spintronics[3,4] magnetism was induced in intrinsically non-magnetic 2D materials[5] via defects,[6,7] topology,[8] and doping.[9,10] Moreover during the recent years a broad range of layered intrinsic magnetic compounds has been investigated in the search of persistent magnetic response at the 2D limit.[11–13] Monolayers of $FePS_3$,[14] $CrI_3$,[15] $Cr_2Ge_2Te_6$,[16] and $Fe_3GeTe_2$[17] are some of the most studied 2D magnetic systems thus far. Interesting enough, the magnetic state of these systems can be controlled by means of electric fields. Magnetoelectric coupling opens the door for the design of 2D magnetic material-based devices for electrically coupled spintronics.[18–20] Further potential ways to control magnetic devices are interface-based spin-transfer torque[21] and voltage controlled magnetic anisotropy.[22] Magnetic anisotropy is the key requisite for perceiving 2D magnetism.[23] Generally, magnetic properties of solid materials depend on the spin orbit coupling and exchange interaction.[24] The reciprocity among exchange interaction, spin orbit coupling, and Zeeman effect describes the orbital moment, magneto crystalline anisotropy, and the ramifications of external magnetic fields.[25] The magnetocrystalline anisotropy and the spin Hamiltonian, being material dependent can be altered by strain engineering[26] and optical tuning.[27,28] Furthermore, tuning the twist angle between bilayers of a 2D homo system enables the possibility to alter the interlayer exchange interaction and controlling the spin degree of freedom in 2D magnetic materials.[29]

For device applications, 2D magnetic materials enable atomically sharp interfaces and the preservation of long-range magnetic ordering down to their mono-layers.[30] However, ambient stability is a limiting factor for their integration into future technologies. One of the possible approaches to overcome this hurdle, is introducing small quantities of V,[31] Fe,[32] or Co[33,34] in transition metal dichalcogenide monolayers via substitution of the transition metal ions. Such type of diluted magnetic semiconductor mono-layers are air stable and exhibit ferromagnetic ordering at room temperature.[31–35] In these systems, controlled dopant concentration and their dilution in the host 2D semiconductor matrix are the key factors.[36]

In contrast to the above-mentioned synthetic 2D systems, there is a variety of vdW minerals[37–41] naturally occurring bulk layered crystals which offer a wide range of structural and compositional variations.[42–47] However, besides natural graphite and molybdenite most of the other explored 2D materials are obtained from synthetic sources, while the properties of naturally occurring vdW species remain mostly unexplored.[37,48] One of the first natural 2D phyllosilicates to be employed in electronics and nanomechanics was talc.[49–52] Recently, mica group members started attracting attention as multifunctional insulators in 2D electronic applications,[42,53,54] and especially in novel memory concepts with 2D semiconductors.[55,56] As for the magnetic vdW minerals, the naturally occurring sulfosalt cylindrite with vdW superlattice was explored.[57] While it is possible to exfoliate monolayers, the bulk antiferromagnetic Néel temperature ($T_N$) was found to be below 20 K.[57] Furthermore, a recent study has demonstrated the possibility to exchange interlayer ions in micas and vermiculites and to fabricate twistronic superlattices, causing interlayer species to reorder accordingly with the twisting angle via the stacking method.[58] Bulk iron (Fe)-rich phyllosilicates – such as minnesotaite and annite – were reported to have an in-plane easy axis, commonly exhibiting layered antiferromagnetism.[59] Thus far, these materials were not explored at the 2D limit. Furthermore, synthetic crystals of flurophlogopite mica were demonstrated to exhibit strong paramagnetic response at low temperatures.[60] We have reported the first observation of weak ferromagnetism in naturally occurring Fe-rich talc at room temperature, confirming that the magnetic response remains down to the monolayer, and that even monolayers are fully air stable.[61]

In this article, we explore minnesotaite, annite, and biotite, as iron-rich members of the phyllosilicate family. These layered systems are inherently magnetic, fully stable under ambient conditions, and can be thinned down to ultra-thin films, and monolayers with the same mechanical cleavage methods as applied for the other 2D materials.[1] They integrate local moment bearing ions of Fe via magnesium (Mg) or aluminium (Al) substitution in their octahedral sites, with substitution ratios of up to 100% with respect to the iron-free species of talc and phlogopite. Due to self-inherent capping by silicate/aluminate tetrahedral groups, the monolayers are air-stable. These minerals exhibit strong paramagnetic behaviour at room temperature and order ferrimagnetically or antiferromagnetically below the critical ordering temperature. We provide correlations between the iron concentration, layer structure, iron oxidation states, and their magnetic response. Furthermore, we discuss the possibilities to tune the structure of magnetic phyllosilicates in order to achieve higher critical ordering temperatures. Our study of 2D materials obtained by mechanical exfoliation of naturally occurring layered magnetic materials may initiate the development for controllable synthesis of the magnetic phyllosilicates with tailored properties.

## 2 Results and discussion

### 2.1. Mineral specimens, layer structure, and exfoliated ultra-thin films of iron-rich phyllosilicates.

Phyllosilicates of interest belong to the class of 2:1 layered minerals composed of two-dimensional sheets of M-(O,OH) octahedra sandwiched between two inward pointing sheets of linked T-O tetrahedra. Cations in the M site are mostly Mg, Al and Fe, while the T sites are occupied by Si and Al.(see Figures 1a-d). They are located at the central octahedral sheets with connection to four O and two adjacent OH groups. In talc, the main M cation is Mg. In some cases, the octahedral sites are also occupied by Fe, cobalt (Co), or nickel (Ni) via cationic substitution,[62] *e.g.* for Fe-substituted talc, the substitution ratio (η) can be expressed as $η = N_{Fe}/(N_{Fe} + N_{Mg})$, considering that the only option for the Fe incorporation is the substitution of the central Mg ion. Two major differences of micas in comparison to talc group minerals are the presence of $AlO_4$ tetrahedral groups in the silicate double layer (with the ratio of Al to Si atoms being close to 1:3) and of larger ionic radius interlayer species (mainly K,Na) located between the 2:1 sheets. The role of the latter is to keep the structure stoichiometry neutral by donating one electron and compensating for the electron deficit created by $Al^{3+}$ ions replacing $Si^{4+}$ ions.[46,63]

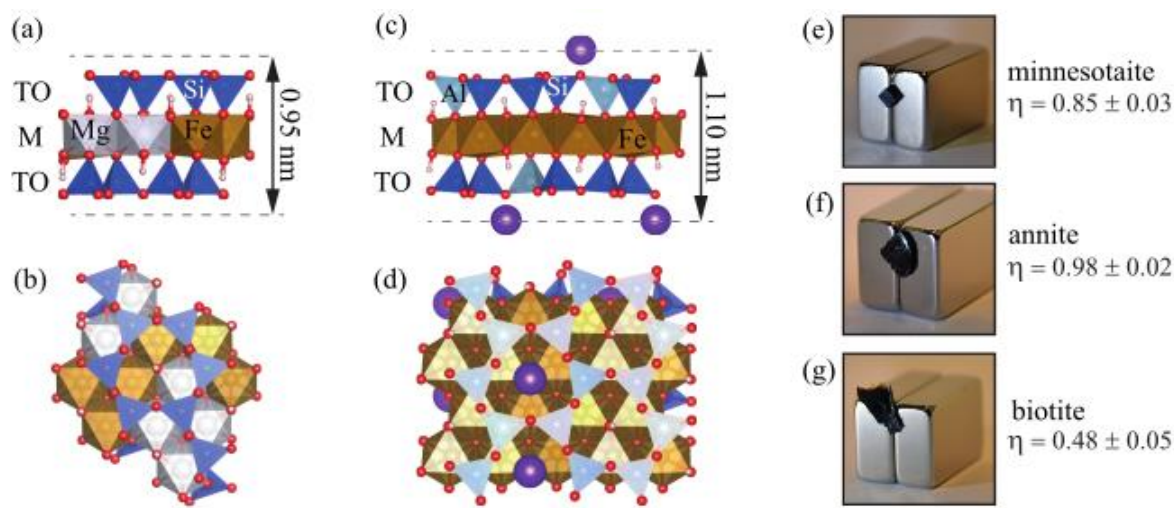

**Figure 1:** Side (a,c) and top (b,d) views of Fe-substituted talc - minnesotaite (a,b), and Fe-substituted mica – annite (c,d). The structure of biotite is not shown as it resembles the annite structure with lower Fe concentration. Presented are relaxed structures obtained from *ab initio* calculations. The octahedral units with substituted Fe are indicated by orange colour, and Mg containing octahedrals are shown in grey. The interlayer species (K) are represented as purple spheres, Si and Al tetrahedral groups are shown in dark and light blue, respectively. The H and O atoms are denoted as white and red spheres. (e-g) photographs of the examined mineral specimens of minnesotaite, annite, and biotite, with their Fe substitution ratio ($\eta$) indicated. The minerals are clinging to a side between two permanent magnets (height of the magnets ~2 cm). Note that as they are (super) paramagnetic at room temperature, the specimens are attracted to the strongest field gradient (*i.e.* between the two opposing poled magnets).

Dark coloured mica is commonly referred to as biotite by geologists. Mineralogically, it corresponds to a group of dark micas including the species (endmembers) phlogopite, siderophyllite, annite and eastonite. Phlogopite and annite do not incorporate Al in the M site. In eastonite and siderophyllite additional Al substitutes for (Mg, Fe) balanced by Al for Si in tetrahedral sites (Tschermak's substitution). In most natural biotites there is considerable substitution of this type.[64]

The structural characteristics of talc group minerals and micas are mainly determined by the composition of the octahedral sheet and the fraction of $Al^{3+}$ tetrahedra. For simplicity, we assume that talc contains no Al at all, while the fraction of $Al^{3+}$ ions in trioctahedral micas like annite corresponds to exactly 1/4 of the total number of the tetrahedral groups.[65] In this case, the main variables relevant for talc are $\eta$ (the fraction of octahedral sites occupied by Fe), and an about 0.15 nm smaller layer separation in the case of talc members due to the lack of the interlayer ions.[49,66] Considering the relaxed structures of both Fe-rich talc and mica, Fe ions should be in a valence state of $Fe^{2+}$, as this is energetically more favoured over the $Fe^{3+}$ state.[61,65] However, in the particular case of minnesotaite, rather rigid tetrahedral silicate groups do not allow for strain relaxation upon the substitution of Fe in the central site. Consequently, a complex modulated layer structure was proposed to occur.[67] This is supported by our observations of the typical flake morphologies, as mechanical exfoliation yielded elongated multilayer flakes and in general a much lower degree of cleavage than with the samples of annite and biotite. In the case of micas, a fraction of the aluminate tetrahedral groups allows for more flexibility and the monolayers can have up to 100 % of the central site substitution without compromising the layered structure. In general, structural defects, vacancies, and reconstructions due to strain can promote a significant fraction of $Fe^{3+}$ ions in the central octahedral site.[68]

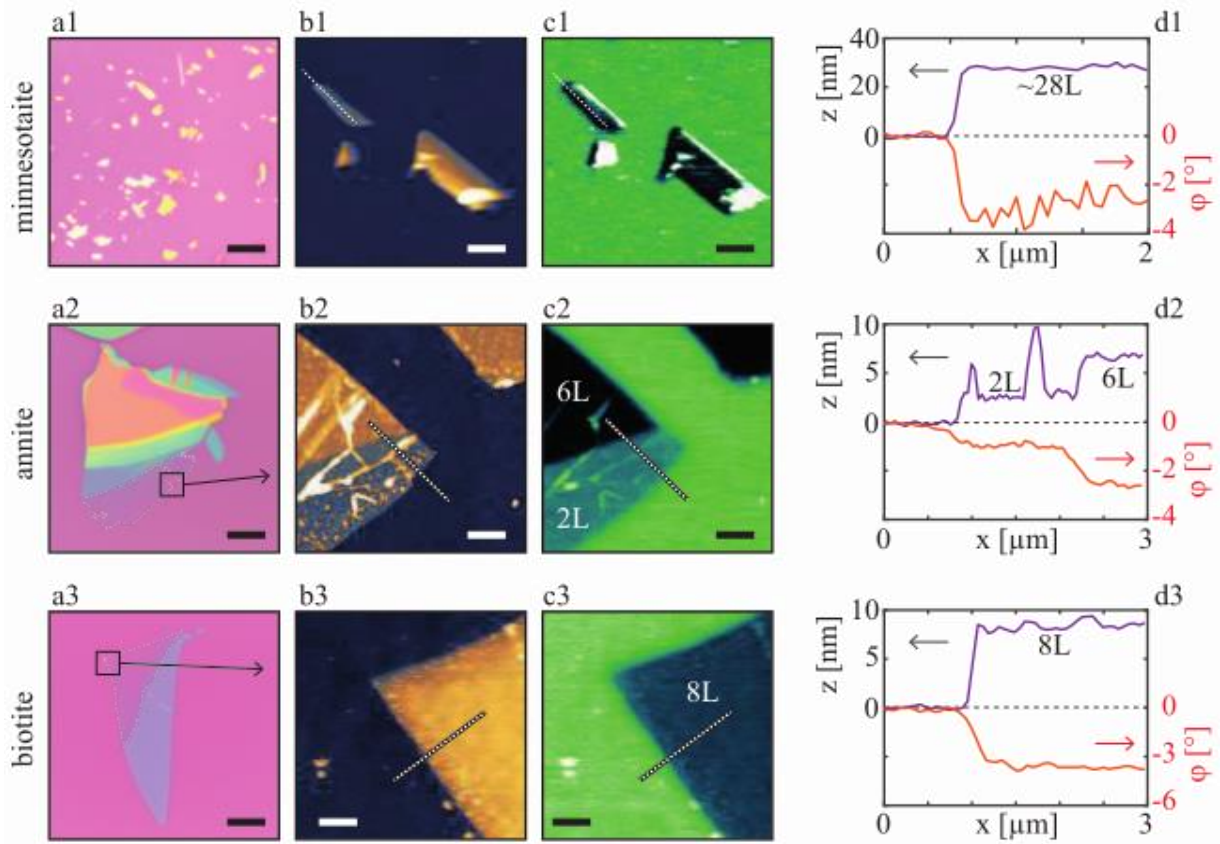

**Figure 2:** (a1-a3) optical micrographs (60×60 μm$^2$, scale bar 10 μm) of exfoliated minnesotaite, annite, and biotite flakes, respectively. Dashed lines in (a2) and (a3) highlight the thin flake regions and the solid squares mark the areas from which the AFM micrographs are presented. (b1-b3) AFM topography images of the exfoliated flakes (6×6 μm$^2$, scale bar 1 μm, z-scales 100 nm, 15 nm, and 15 nm respectively). (c1-c3) second-pass phase lag in 180 mT out-of-plane external field, of the same flake areas as in (b1-b3). (d1-d3) presents flake height (left y scale) and MFM phase lag (right y scale) cross-sections, taken perpendicular to the flake's edges. Single lines are presented without line averaging. The cross-sections are indicated in (b1-b3 and c1-c3) by dotted white lines.

Figures 1a-d depict the 3D representation (side and top views of the relaxed structures obtained by *ab initio* calculations) of Fe-substituted talc and mica monolayers. Photographs of the three mineral specimens used in the study (sticking to the side of the permanent magnets) are shown in Figure 1e-g. The chemical composition of all the samples was determined using quantitative electron probe micro analysis -wavelength dispersive X-ray spectroscopy (EPMA-WDS) and X-ray photoelectron spectroscopy (XPS). The examined minnesotaite crystals were obtained from the mineral collection of the University of Minnesota (sample number 14508). The mineral aggregate consists of small crystalline grains of minnesotaite (up to 100 μm in size) with occasional pure talc flakes and $SiO_2$ inclusions. Based on the systematic Raman spectra taken from the mineral and numerous WDS measurements, we estimate that above 95 % of the grains belong to minnesotaite. The examined minnesotaite samples have $\eta = (0.85 \pm 0.03)$. The empirical formula calculated from EPMA-WDS analyses on the basis of 11 oxygen equivalents per formula unit is $(Fe_{2.5}Mg_{0.5})Si_{4.4}O_{10}(OH)_2$. The detected slight excess of Si could be related to $SiO_2$ inclusions between minnesotaite grains and to residues of the poly-dimethylsiloxane (PDMS) films used in sample preparation for the WDS experiments (see also methods section).

The examined specimen of annite was obtained from the mineral collection of the Royal Ontario Museum (specimen number M42126.1). The bulk sample is a single-crystal (about 8 mm in diameter). Inclusions of other species were not detected. The empirical formula, determined by WDS, is: $K_{1.02}(Mg_{0.05}Fe_{2.94})Al_{0.94}Si_{3.28}O_{10}(OH)_2$. Iron was found to occupy almost 100 % of the central octahedral sites; $\eta = (0.98 \pm 0.02)$. In both cases of annite and biotite samples, the detected amount of Si (about 6-8 %) can be assigned to PDMS residues, which have been also observed on the crystal graphite support used in WDS experiments.

The examined sample of biotite was obtained from the mineral collection of the Chair of Resource Minerology at the Montanuniversität Leoben (sample number: 9127). The mineral is a single-crystal (about 10 mm in diameter), and inclusions of other species were not detected. The empirical formula (assuming no central octahedral ion vacancies) is: $K_{0.82}(Mg_{1.02}Fe_{1.42}Al_{0.54})(Al_{0.96}Si_{3.23})O_{10}(OH)_2$, as determined by WDS. In the specimen, $\eta = (0.48 \pm 0.05)$ was found. Considering that only Fe and Mg occupy the central octahedral site, the Al amount would be in excess. Therefore, Al ions are likely to occupy about 18 % of the octahedral sites.[69] The observed deficit in the interlayer K ions could lead to incorporation of about 15 % of $Fe^{3+}$ with respect to the total number of Fe ions. However, as seen later in the text, this is not sufficient to explain the observed $Fe^{2+}/Fe^{3+}$ ratio, indicating a more complex relation between the iron oxidation state and the structural defects.

Monolayer and multilayer 2D flakes of minnesotaite, annite, and biotite were prepared from the bulk mineral specimens via micromechanical exfoliation[1] and transferred primarily onto $Si/SiO_2$ chips (300 nm oxide layer), see Figure 2(a1-a3). All these vdW materials are dielectrics with a band gap of approximately 5-6 eV.[56] Hence, the exfoliated flakes have the issues related to the optical contrast in the visible similar to hexagonal boron nitride.[44,49,70–72] Owing to the high-crystallinity of the starting mineral specimens, high-quality annite and biotite flakes were obtained with large uniform areas (over 30 μm in diameter), and terraces with thicknesses down to few layers. For minnesotaite flakes, due to polycrystallinity of the bulk that likely results from the reconstruction of the crystal structure due to high Fe substitution, cleavage is less pronounced and predominantly elongated multi-layer flakes in form of sticks were observed,[67] as shown in Figure 2(a1-c1). Typical morphologies of the thin flakes of annite and biotite are depicted by the atomic force microscopy (AFM) topography images in Figure 2(b1-b3).

To map the local magnetic response of the flakes, two-pass magnetic force microscopy (MFM) measurements were carried out in the presence of an externally applied out-of-plane magnetic field of 180 mT. Figures 2(c1-c3) present the MFM second pass phase lag images that correspond to the in-field local magnetic response at room temperature. As the samples are superparamagnetic/paramagnetic at room temperature, the external out-of-plane magnetic field is stronger above the flakes in comparison to the surrounding weakly diamagnetic $Si/SiO_2$ support. Consequently, the magnetized probe is pulled stronger to the surface, yielding a negative contrast with respect to the substrate.[61] A comparison between the topography and the MFM phase lag is provided by their cross-sections in Figure 2(d1-d3). A correlation between the height and the second pass phase lag were observed, as illustrated for annite where a bi-layer (2L) to six-layer (6L) flake step is presented in Figure 2(b2-d2). The magnetic behaviour is strongly dependent on the amount of Fe ions substituting Mg ions in the octahedral sheet. Moreover, since all other bonds are saturated, Fe ions can take on both 2+ and 3+ valence states as a result of structural variations. As will be shown below, the fraction of $Fe^{3+}$ ions can be significant and their presence makes a strong impact on magnetic properties of phyllosilicates.

To verify the structure and to confirm that the crystallinity is preserved upon mechanical exfoliation, thin flakes of minnesotaite, annite, and biotite were measured by micro-Raman spectroscopy, and their spectral features were analysed with respect to Fe incorporation into the phyllosilicate matrices. For these measurements, thin flakes (between 20 nm and 200 nm) were transferred onto the surface of highly oriented pyrolytic graphite (HOPG).[73] HOPG as a substrate provides good heat dissipation and allows for the long accumulation time needed in order to resolve the spectral features of ultra-thin phyllosilicate flakes. In addition, the vibrational modes of HOPG do not overlap with neither low- nor high-frequency modes of the examined phyllosilicates, which is not the case for $SiO_2/Si$ support. Figure 3 represents the Raman spectra with fundamental and OH associated vibrations for each of the phyllosilicates along with the optical images of the measured thin flakes.

Mostly Raman features of phyllosilicates are predominantly observed in the two spectral ranges: from 250-1200 cm$^{-1}$ and 3000-3800 cm$^{-1}$. The Raman peaks in the spectral range below 600 cm$^{-1}$ are an indication of the complex set of translational motion of cations in the octahedral site, followed by the peaks between 600-800 cm$^{-1}$ related to the fundamental vibrations $(Si – O_b – Si)$.[74] The Raman peaks in the spectral range of 800-1150 cm$^{-1}$ results from the stretching mode of Si and non-bridging oxygen.[74] The peaks from the stretching mode of OH groups in the octahedral sites and potential intercalated $H_2O$ can be observed in the spectral range of 3000-3800 cm$^{-1}$. In phyllosilicates, the peak positions of the characteristic OH vibrations depend on the difference in the effective ionic radii among $Fe^{2+}$ (0.78 Å), $Mg^{2+}$ (0.72 Å), $Fe^{3+}$ (0.645 Å), and $Al^{3+}$ (0.53 Å).

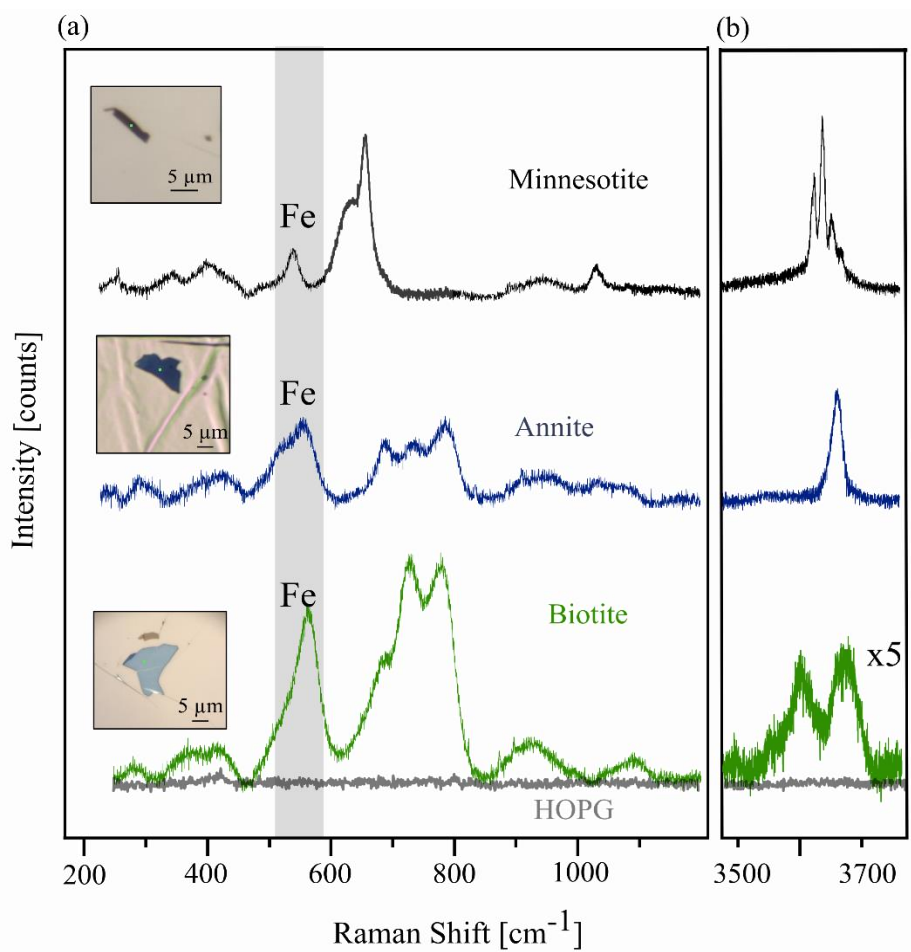

**Figure 3:** Raman spectra of thin Fe-rich phyllosilicates, showing the fundamental vibrations (a) (250-1200 cm$^{-1}$ range), and the OH vibrations (b) (3500-3700 cm$^{-1}$ range). Top-down, the spectra correspond to minnesotaite, annite, and biotite respectively. Insets present optical micrographs of the flakes on HOPG, from which the measurements were taken, and the green dot indicates the laser positioning. The background spectrum of graphite is shown together with the spectrum of biotite.

Different spectral changes were observed in the case of minnesotaite, annite and biotite, depending on their structural difference and the variation of Fe concentration in the mineral. Figure 3a shows the Raman spectra of fundamental vibrations for all three minerals in the spectral region: 200-1150 cm$^{-1}$. Raman spectra exhibit the presence of Si-O-Si vibration modes followed by the Fe-O vibration modes in the spectral range of 250-500 cm$^{-1}$ and 500-600 cm$^{-1}$ respectively. In the case of annite and biotite the triplet of peaks between 700-820 cm$^{-1}$ is assigned to the octahedral-tetrahedral bridging oxygen vibration mode,[74] which is observed in the Raman spectra of minnesotaite as a doublet at lower frequencies (600-720 cm$^{-1}$), as a result of the absence of Al in the minnesotaite structure.

Figure 3b shows the Raman spectra of minnesotaite, annite and biotite in the hydroxyl stretching region. In comparison to talc,[61,74] minnesotaite exhibits major spectral differences due to the substitution of Fe in the octahedral sites. The downshift of peak positions are attributed to the mass effect.[74] A complex OH feature was observed with the three major peaks at 3655, 3626 and 3639 cm$^{-1}$. These different vibrational modes relate to the presence of Fe$^{2+}$, Fe$^{3+}$ and Mg$^{2+}$ in the octahedral sites forming [Fe$^{2+/3+}$O$_4$(OH)$_2$] and [Mg$^{2+}$O$_4$(OH)$_2$]. In the case of annite and biotite, a strong peak at 3670 cm$^{-1}$ and a doublet peak at 3640 and 3680 cm$^{-1}$ was observed respectively. The splitting in the peaks of biotite is related to the diverse occupancy of Fe, Al and Mg ions at octahedral sites,[75] which is further supported by the presence of single peak for annite, containing 100 % Fe in the octahedral site. Furthermore, in the case of biotite the broadening of OH modes suggest a significant degree of disorder present in the cation sublattice of the octahedral sheet. As discussed below, the presence of the aluminate tetrahedral groups enables both biotite and annite to incorporate higher fractions of Fe$^{2+}$ and Fe$^{3+}$ ions while maintaining the same structure. This is not the case for the talc family members such as minnesotaite.

Therefore, the difference between $Fe^{2+}$ and $Fe^{3+}$ ions in the octahedral groups could play a more significant role in their Raman spectra.

## 2.2 Long range magnetic ordering

To probe the long-range magnetic ordering, superconducting quantum interference device vibrating sample magnetometer (SQUID-VSM) measurements were carried out. In the case of annite and biotite, single crystals were probed and the external magnetic field was applied perpendicular with respect to their basal planes. In the case of minnesotaite, a polycrystalline sample was measured and therefore the grains had random orientation with respect to the externally applied fields, effectively integrating over both in- and out-of-plane components.

Field cooling (FC) and zero field cooling (ZFC) M(T) curves were recorded starting from 400 K to 2 K. The critical ordering temperatures were determined from the global minima of the FC measurements first derivate dM(T)/dT. Figure 4 summarizes the VSM-SQUID results. The M(T) curves indicate the transition from (super)paramagnetic to ferrimagnetic in case of minnesotaite and annite, and anti-ferromagnetic in case of biotite. The transition temperatures considering the dM/dT curves and are 38.7 K for minnesotaite, 36.1 K for annite, and 4.9 K for biotite, as presented in the insets of Figures 4(c,f,i).

Magnetization loops M(H) were measured with the external field strength varied between ±6 T, at room temperature and below the critical ordering temperature. The results are presented in Figure 4a,d,g for minnesotaite, annite, and biotite, respectively. In contrast to annite and biotite, which are paramagnetic at 300 K, minnesotaite exhibits a clear saturation of the magnetization, a signature of superparamagnetism. A similar behaviour was observed in the case of iron-rich talc.[61] The field dependent magnetization for minnesotaite at low temperatures displays the ferromagnetic M(H) loop and reaches saturation at around 1.5 T (see Fig. 4a). For minnesotaite, the saturation magnetization of (0.7 ± 0.4) emu/g at 10 K was observed with a coercivity of 200 mT which is in good agreement with the values reported in the literature.[76]

For annite and biotite samples, the temperature dependent magnetic susceptibility ($\chi(T)$) was estimated from two FC M(T) curves with different applied external fields (see methods for details). The inverse of the estimated $\chi(T)$ in the paramagnetic regime is presented along with FC and ZFC M(T) curves in Figure 4e,h. The Curie-Weiss fit of the susceptibility in the high-temperature range provides a consistent estimate of the local moment per Fe ion between 4 and 5 $\mu_B$, suggesting that both compounds contain mixtures of $Fe^{2+}$ and $Fe^{3+}$ ions. A more detailed fit was performed by assuming that the paramagnetic response originates from a mixture of $Fe^{2+}/Fe^{3+}$ ions, which is expected in sheet silicates.[77] The fit can be described by the expression:

$$\chi(T) = \chi_0 + \frac{(1-p)C_1 + pC_2}{T-\theta} \qquad (1)$$

with $C_1 = \eta_{Fe}\mu_B^2 g_1^2 S_1(S_1+1)/3\,k_B$, $C_2 = \eta_{Fe}\mu_B^2 g_2^2 S_2(S_2+1)/3\,k_B$, where $g_1, S_1$ and $g_2, S_2$ correspond to the Landé factor and spin quantum number of $Fe^{2+}$ and $Fe^{3+}$, respectively; $\eta_{Fe}$ denotes the average number of the octahedral sites occupied by Fe ions; $\theta$ is the paramagnetic Curie temperature for $Fe^{3+}$ ions; $\chi_0$ is the temperature independent susceptibility (containing the background and the van Vleck paramagnetism) and $p$ represent the fraction of $Fe^{3+}$ relative to the total amount of iron . Setting $g_1 = g_2 = 2$ (which is a reasonable assumption at least for the annite and biotite samples because the field is orthogonal to the easy plane of magnetization), $S_1 = 2$, $S_2 = 5/2$, partial Curie constants are found to be: $C_1 = 9.54 \cdot \eta_{Fe}$ K emu mol$^{-1}$, $C_2 = 13.92 \cdot \eta_{Fe}$ K emu mol$^{-1}$.

Fitting the susceptibility above the ordering temperature to the equation (1), we obtain a $Fe^{2+}/Fe^{3+}$ ratio of 78/22 for annite and 53/47 for biotite, which is consistent with previous studies of Fe-rich sheet silicates.[77,78] In the case of minnesotaite, a more complex temperature dependence of the M(T) curves was observed (see Fig. 4b).

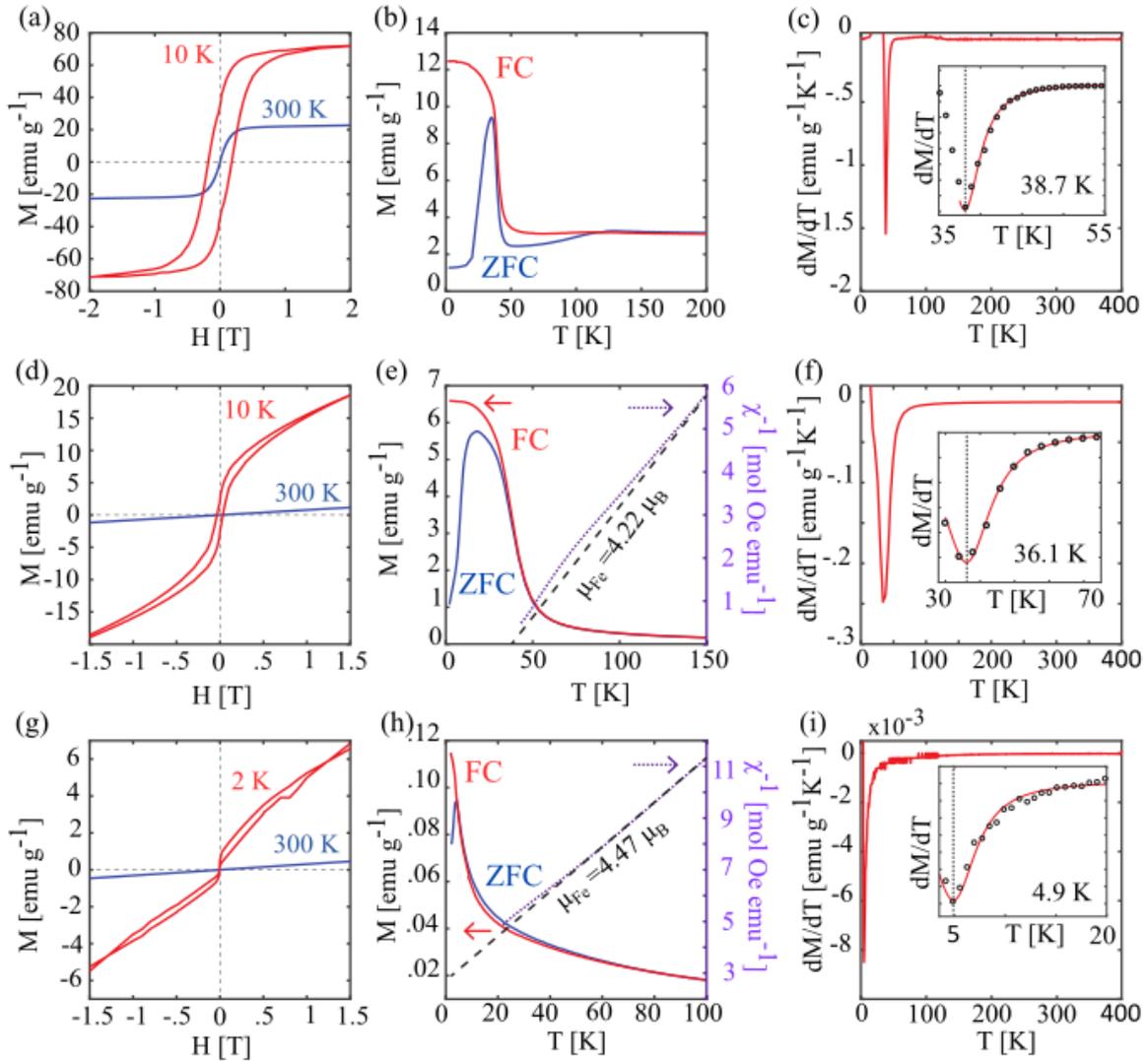

**Figure 4:** (a,b) M(H) loop and FC-ZFC M(T) curves for minnesotaite. (d,e) M(H) loop and FC-ZFC M(T) curves for annite, and (g,h) for biotite. All magnetic moments are scaled with respect to the mass of the measured samples. (c,f,i) dM/dT curves for minnesotaite, annite, and biotite, respectively, with insets focusing on the global minima and denoting the corresponding critical ordering temperatures. Mind that in the M(T) curves (b,e,h) the temperature scales are not the same, but rather focus on the regions near the transitions.

The paramagnetic Curie temperature, $\theta = 40$ K, for annite reveals the dominance of ferromagnetic interactions between Fe ions. However, for biotite $\theta = -27$ K was obtained, highlighting predominantly antiferromagnetic interactions and also a large degree of frustration, as $\theta/T_c \gg 1$. This behaviour can be attributed to the Fe/Mg disorder on the octahedral sublattice. Overall, the trends in the magnetic ordering tendency are consistent with the composition of the samples.

### 2.3 Incorporation of Fe in the phyllosilicates

To probe the oxidation state of iron in the phyllosilicate matrices, X-ray photoelectron spectroscopy (XPS) and X-ray absorption spectroscopy (XAS) were performed. The XPS/XAS spectra of iron compounds considerably differs depending on the ion oxidation state and the spin configuration.[79,80] In particular, the oxidation states of iron can be distinguished as $Fe^{3+}$ (always high spin) and $Fe^{2+}$ (can be high spin and low spin) by the binding energies of the Fe2p core-level peaks and their satellites in the spectra. According to the theory, high-spin $Fe^{2+}$ and $Fe^{3+}$ configurations are multiplets due to spin-orbit and crystal field interactions, which can be described by the Gupta and Sen (GS) model.[81] Satellite peaks are generally characteristic of high-spin configurations due to the energy loss of photoelectrons on spin interactions.[82] Iron states in silicates are analysed by a simplified model, considering core levels and their satellites as singlet peaks with binding energies for $Fe^{3+}$ typically 2-3 eV higher than for $Fe^{2+}$.[83]

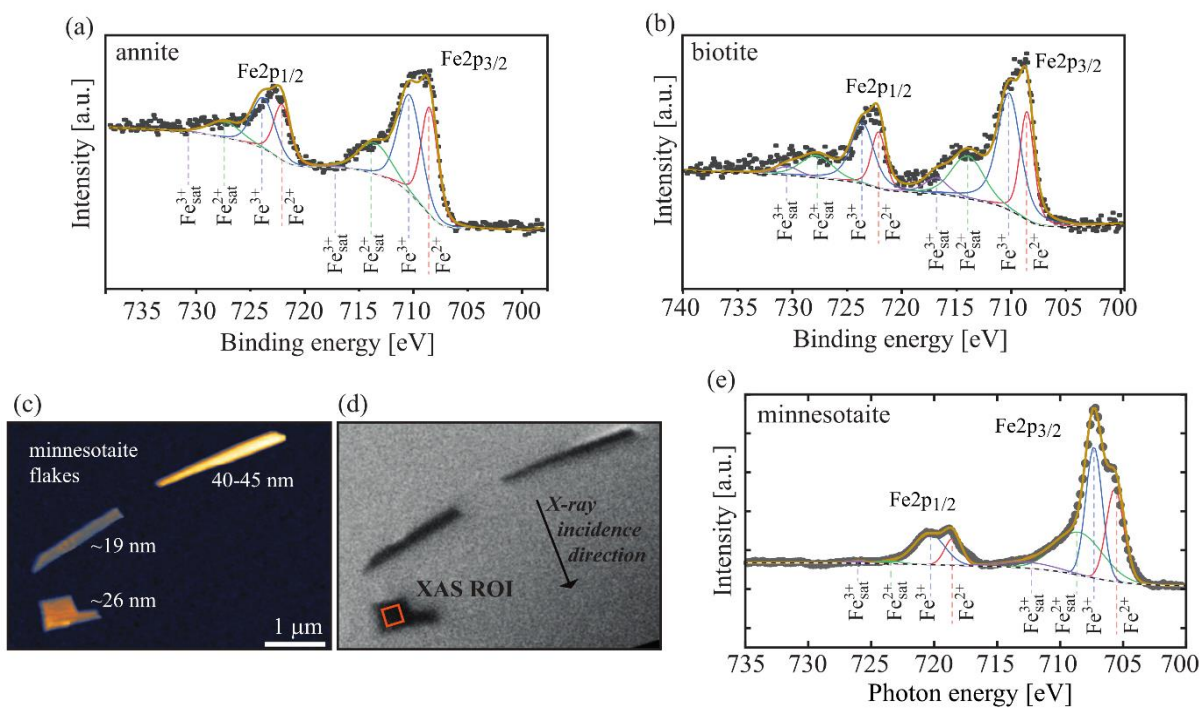

**Figure 5:** (a,b) XPS spectra of annite and biotite, (c) AFM image of minnesotaite flakes with their thickness (d) PEEM Image of minnesotaite flakes obtained during XAS measurements across the Fe $L_{3,2}$-edge. Squared orange box indicates the region from which the XAS spectrum represented in (e) has been obtained for minnesotaite.

XPS was applied to annite and biotite single crystals as large-enough (about 1 mm in diameter) crystalline samples with no inclusions were available (see methods for the details on sample preparation for XPS). The Fe2p spectra of annite and biotite are shown in Figure 5(a,b). Considering the simplified model, annite possesses $Fe^{2+}$ in high spin configuration which is perceptible from the satellite peak of the correspondent multiplet. However, high-spin $Fe^{3+}$ is not so unambiguous to deconvolute, since the $Fe^{3+}$ satellites either have very low intensities or are hidden under the Fe2p$_{1/2}$ peak. Similar to annite, in the case of biotite a mixture of $Fe^{2+}$ and $Fe^{3+}$ states in high spin configuration was observed. According to the model, the $Fe^{2+}/Fe^{3+}$ ratio is 45/55 and 35/65 for annite and biotite, respectively. The values obtained reveal a larger fraction of $Fe^{3+}$ ions than what is obtained from the Currie-Weiss fit of the inverse $\chi(T)$, although exactly the same crystals used for the SQUID magnetometry measurements were also used to record the XPS spectra. The magnetization measurements probe the entire volume of the sample, while the photoelectrons are obtained from the surface. The observed discrepancy in the $Fe^{2+}/Fe^{3+}$ ratio can be introduced by a possible surface-related defects. However, both techniques reveal a mixture of $Fe^{2+}$ and $Fe^{3}$ ions, while in the perfectly ordered lattice only the $Fe^{2+}$ should be present as suggested by the *ab inito* calculations.

Considering the polycrystallinity of minnesotaite, in order to avoid the influence of potential contaminants XAS imaging across the Fe $L_{3,2}$-edge for minnesotaite was measured on exfoliated single crystal flakes by means of x-ray photoemission electron microscopy (PEEM). The measurements were carried out on multilayer 2D flakes of minnesotaite transferred onto gold-coated Si chips. Upon illumination of the samples, the absorbed fraction of the X-rays leads to the excitation of inner shell electrons and a cascade of secondary electrons, which are measured by the microscope as a function of the incoming X-ray photon energy. The absorption spectra indicate the partial occupation of $Fe^{2+}$ in a high-spin configuration, which is evident from the satellite peaks. The same analysis procedure was applied as for the XPS spectra of annite and biotite, estimating $Fe^{2+}/Fe^{3+}$ ratio for minnesotaite of 54/46.

**Discussion**

Phyllosilicates provide a natural solution of ambient stability for 2D intrinsic magnetism and can thus potentially represent a robust platform for future technological advancements and next generation data storage devices. Here we demonstrated that the magnetic ordering in naturally occurring iron rich phyllosilicates showing 2:1 layering (minnesotaite, annite, and biotite) persists down to few monolayers. An important question that remains to be answered is how to tune the critical temperature and whether it is possible to significantly increase it.

A trend can be drawn from our results in conjunction with previously published data on bulk crystals. Figure 6 shows the magnetic ordering (the Curie temperature), measured for several classes of Fe-rich phyllosilicates, all having the same octahedral sheet as the main functional unit.[59,76,77,84,85] The magnetic behaviour of all these compounds is similar and can be rationalized in terms of the local electronic structure of $Fe^{2+}$ and $Fe^{3+}$ ions.[77,85] Specifically, the ground state magnetic structure can be described as ferromagnetically ordered layers of in-plane oriented magnetic moment of iron, with dipolar interactions in weakly coupling adjacent layers antiferromagnetically in a perfectly ordered case. Reducing the fraction of Fe or increasing the ratio of $Fe^{3+}/Fe^{2+}$ ions results in stronger disorder within the planes, which eventually destroys ordering and silicates with low Fe content exhibit a complex temperature dependence of the susceptibility akin to that of spin glasses.[86] An important point of this picture is that it suggests the ordering temperature of Fe-rich silicates is determined practically only by exchange interactions between iron ions within individual layers. The effective strength of the exchange interaction is directly related to the paramagnetic Curie temperature, $\theta$, extracted from the temperature dependent susceptibility. Thus, by examining values of $\theta$ we can understand the magnetic behaviour of individual octahedral sheets themselves.

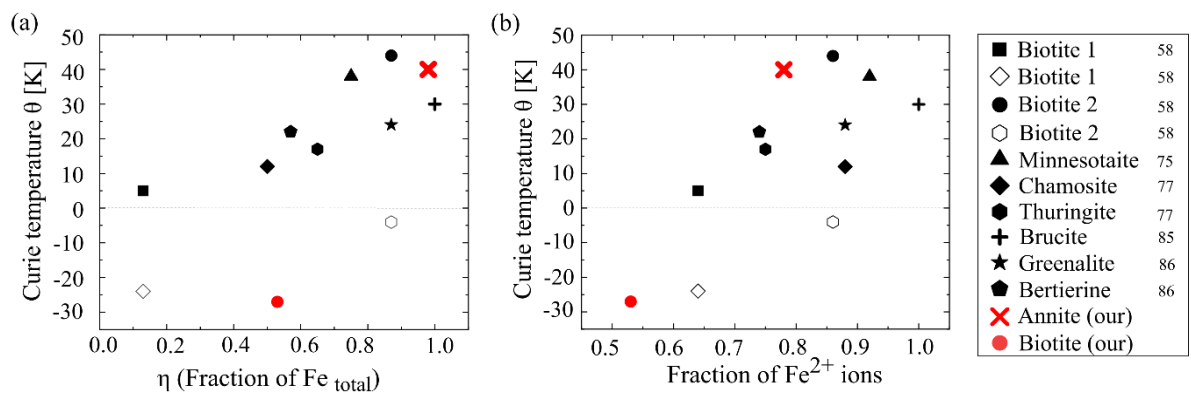

**Figure 6:** The paramagnetic Curie temperature, $\theta$, as a function of (a) fraction of total Fe ions and (b) fraction of Fe-occupied sites by $Fe^{2+}$ ions, in all cases considering the central sites of the octahedral sheet for various phyllosilicates. Most of the data is presented for polycrystalline samples, for which only an effective $\theta$ can be extracted. For monocrystalline samples (Ref. [58] and this work) filled symbols correspond to $\theta_\perp$, where the susceptibility is measured for the magnetic field parallel to the *c*-axis, while empty symbols correspond to $\theta_\parallel$ for the field orthogonal to the *c*-axis.

Considering Figure 6, $\theta$ depends on the total fraction of Fe ions in the octahedral sheets (Fig. 6(a)). When considering the samples with dominant ferromagnetic interactions ($\theta > 0$), simply a higher amount of Fe substitution leads to higher $\theta$. Supposing that there are only $Fe^{2+}$ ions we should expect that the maximum of $\theta$ is observed at the maximum fraction of iron ions equal to 1. However, summarizing results on magnetic measurements for various phyllosilicates with respect also to the Fe oxidation states (Fig. 6(b)), we see that this is not the case. Despite a certain scatter of results (caused by differences in samples, measurements, possible impurities, etc.), there is a clear trend that the Curie temperature increases with increasing fraction of $Fe^{2+}$. However, the observed maximum is located around 75-90 %. It is important to keep in mind that the remaining 10-25 % of metallic ions are not just Mg or Al but these sites are occupied by $Fe^{3+}$ ions. The effect of adding $Fe^{3+}$ ions to a primarily $Fe^{2+}$-occupied sublattice is two-fold. On the one hand, $Fe^{3+}$ ions, with their $d^5$ configuration do not contribute to the magnetocrystalline anisotropy, thus reducing the tendency to 2D magnetic ordering. On the other hand, $Fe^{3+}$ introduces two new types of exchange interaction types, namely $Fe^{3+}$-$Fe^{2+}$ and $Fe^{3+}$-$Fe^{3+}$ interactions. We have estimated NN interactions by performing total energy calculations in an idealized Fe-brucite-like structure (see Methods) and obtained the following values of the interactions (positive meaning ferromagnetic): $J_{2+2+} = 5.7$ meV, $J_{3+2+} \approx 6 J_{2+2+}$, and $J_{3+3+} = -0.5 J_{2+2+}$. These interactions are consistent with the Goodenough-Kanamori rules and show that $Fe^{3+}$ actually enhances ferromagnetic interactions when it is surrounded only by $Fe^{2+}$ ions.[87] This implies that a small amount of $Fe^{3+}$ distributed randomly over the Fe sublattice can increase the value of $\theta$, which could be the origin behind the position of the maximum $\theta$ as a function of $Fe^{2+}$ ion fraction. Note that the higher fraction of $Fe^{3+}$ ions will produce more and more $Fe^{3+}$-$Fe^{3+}$ pairs with anti-ferromagnetic type of interactions, which will result in a reduction of $\theta$ and magnetic frustration, which can also explain the spin-glass-like behavior of phyllosilicates with lower Fe content.

## Conclusions

Our study addresses the relation between the structure and the magnetic properties of iron-rich phyllosilicates, magnetic minerals that can be thinned down to their monolayers using a mechanical exfoliation process. We studied three mineral specimens, the rare talc-like mineral minnesotaite, and the micas annite, as well as more commonly occuring biotite. All three species can be exfoliated down to few monolayers, and their thin-film samples were found to be fully air stable. However, in the case of minnesotaite, peculiar structural changes can occur, producing complex incommensurate patterns and favouring defect formation.[67]

Magnetic measurements show that all three materials exhibit antiferromagnetic or ferrimagnetic ordering at temperatures below 40 K. The analysis of the temperature dependent susceptibility reveals that annite has a positive value of the paramagnetic Curie temperature, implying the dominance of ferromagnetic exchange interactions. On the other hand, biotite, with similar structure as annite has a lower Fe content and a higher fraction of $Fe^{3+}$ ions, all contributing to a negative Curie temperature. This implies predominantly antiferromagnetic interactions and also a large degree of frustration.

The character of interactions can be traced back to both the overall fraction of Fe in the metallic positions of the octahedral sheet and to the relative fraction of $Fe^{3+}$ ions with respect to $Fe^{2+}$ ions. The effect of $Fe^{3+}$ is two-fold, enhancing ferromagnetic interactions mediated by $Fe^{2+}$-$Fe^{3+}$ pairs, and reducing the crystalline anisotropy and forming $Fe^{3+}$-$Fe^{3+}$ pairs which have a detrimental effect on the in-plane ferromagnetic ordering. This behaviour is supported by *ab-initio* estimates of nearest-neighbour magnetic exchange interactions. This is also supported by our meta-study and previously published results of magnetic phyllosilicates, which show that the highest Curie temperature is not achieved for the highest content of $Fe^{2+}$ ions (achieved in Fe-brucite) but for compounds with about 10-25% of $Fe^{2+}$ ions replaced by randomly distributed $Fe^{3+}$ ions.

## Methods

### Sample preparation

The flakes of minnesotaite, annite and biotite were prepared by micromechanical exfoliation using sticky tape (Nitto Denko ELP BT150ECM). After multiple pealing against two pieces of the tape, the material was deposited on $SiO_2$/Si chips (300 nm oxide layer), gold-coated Si chips, or cleaved graphite crystals, via PDMS. After slowly peeling off the PDMS, flakes with different thicknesses were identified with the help of an optical microscope.

### SQUID magnetometry

Long range magnetic ordering was measure using superconducting quantum interference device - vibrating sample magnetometer (SQUID-VSM) MPMS2 from Quantum Design. Bulk crystals were placed inside a plastic straw holder and inserted into the electromagnet. All the magnetization curves were normalized to the diamagnetism of substrate and the values of magnetic moment was normalized to the mass of the measured crystals. M(T) measurements in the FC case were done with two different externally applied fields. For the minnesotaite and biotite samples external fields of 5 mT and 20 mT were applied while 10 mT and 100 mT were applied for the annite sample. The susceptibility was estimated from the two FC M(T) curves by a simple linear approximation considering the M(H) point of origin as the third and fixed point. The results are consistent with the susceptibility obtained from the derivates of M(H) loops in the paramagnetic regime (at 45 K and 300 K). The scaling of the susceptibility was done considering the molar mass of annite to be 1032.8 g·mol$^{-1}$, and 921.94 g·mol$^{-1}$ for biotite. Curie-Weiss fit of the susceptibility was applied in a 100 – 400 K range for annite and 50 – 400 K range for biotite.

### Structural characterization

Element mapping and quantitative analysis was performed using the Jeol JXA8200 electron probe microanalyzer at the Chair of Resource Mineralogy at Montanuniversität Leoben. Investigations were carried out in WDS mode with counting time ranging from 10 s to 300 s. Other operational parameters include 10-100 nA beam currents and analysing voltages of 10-15 KV. Ka lines of Si, Al, Mg and Fe were used for analysis. ZAF corrections was applied and kaersutite and magnetite were used as standards for calibration.

### X-ray photoelectron spectroscopy

X-ray photoelectron spectrometer ESCALAB 250Xi was used to determine the oxidation state of iron. Bulk crystals were were used in the experiments. Spectra were acquired with monochromated Al Ka excitation (hv = 1486.7 eV) at pass energy of 20 eV providing a spectral resolution of 0.5 eV. No external charge compensation was used, instead the charging effect was corrected to the Si$_{2p}$ peaks (101.6 eV) assuming their chemical states in the studied silicates to be same. Spectra was acquired and processed using the Advantage software.

### X-ray absorption spectroscopy

X-ray Absorption spectroscopy was performed using the scanning photoemission electron microscope at the UE49-PGM SPEEM beamline of BESSY II. 2D flakes of minnesotaite were transferred on gold-coated Si chips. Images and spectra were recorded using linear polarized X-rays at a sample temperature of 40 K.

### Raman spectroscopy

Raman features of the samples were recorded using Horiba LabRam HR evolution confocal Raman spectrometer. Thick flakes (between 20 nm and 200 nm) of the minerals were transferred on HOPG. The measurements were carried out with 532.1 nm laser (1800 gr/mm and 100 mW power on the sample surface), and 100x magnification lens. No sample degradation/damage was noticed after continuous illumination under the laser beam.

### AFM and MFM measurements

Topography analysis (using AFM) and local magnetic moment measurements (using MFM) were carried out on Horiba-AIST-NT Omegascope system at room temperature. For topography analysis, non-ferromagnetic PtIr-coated Electrical Force Microscopy (EFM) probes of type ACCESS-EFM from AppNano (~ 30 nm tip curvature radius, ~ 63 kHz resonance, ~ 2.7 Nm$^{-1}$ force constant) were used. MFM signals were recorded using TipsNano MFM01 probes with CoCr coating (~40 nm tip curvature radius, ~70 kHz resonance, ~5 N/m force constant). Prior to MFM measurements, counter check of probes magnetic response were carried out on a magnetic hard drive (MFM test samples). The probes were subjected to out-of-plane magnetic field with flux density of 200 mT for around 30 minutes prior to MFM experiments. In field measurements were performed on a "home-built" sample holder with permanent magnet of 180 mT, estimated at the sample surface. Hall probe (M-test MK4, Maurer Magnetic AG) was used to confirm the flux density of externally applied magnetic field. MFM signals were recorded in a two-pass regime, with second-pass lift height of 20 nm. MFM phase and AFM topography were mapped and processed via Gwydion v2.55 (an open source software for SPM analysis).

## DFT calculation

DFT calculations were performed using the projector augmented waves[88] method as implemented in the Vienna Ab-initio Simulation Package.[89–91] Generalized gradient approximation with the Perdew-Burke-Ernzerhof parameterization for solid was employed as an exchange-correlation functional, as it was shown to give the best-relaxed structure for talc.[92,93] Furthermore, we use the value $U_{eff} = 4$ eV within the DFT+U rotationally invariant scheme [94] to better describe localized states of Fe ions. Atomic positions were relaxed to the accuracy in forces of 0.01 eVÅ$^{-1}$. To estimate nearest-neighbour magnetic exchange interactions we construct a model system, representing a $4 \times 2\sqrt{3}$ supercell based on the brucite unit cell, with a pair of adjacent metallic sites occupied by Fe. Naturally, this structure contains only Fe2+ ions. Ions with 3+ states are introduced by removing a hydrogen atom next the target site. The magnetic exchange interaction is evaluated as $(E_{\uparrow\downarrow} - E_{\uparrow\uparrow})/2$, corresponding to a Heisenberg-type Hamiltonian $H = -\sum_{<ij>} J \boldsymbol{S}_i \cdot \boldsymbol{S}_j$. Here, $E_{\uparrow\downarrow}$, $E_{\uparrow\uparrow}$ denote energies of, respectively, anti-parallel and parallel spin configurations for the pair of Fe ions. We have also checked the dependence of the obtained interactions on $U_{eff}$ and the degree of trigonal distortion of oxygen octahedra. Both of these factors can significantly affect the absolute values of the interactions but do not change the signs, the ratios of magnitudes remaining similar to the one given in the main text. The values given in the main text correspond to zero trigonal distortion and $U_{eff} = 4$ eV.

## Acknowledgements

This work is supported by the Austrian Science fund (FWF) via START programme (grant no. Y1298-N). The authors acknowledge infrastructural support of the Montanuniversität Leoben (Raman AFM TERS lab). The authors would like to acknowledge the invaluable support and technical facilities provided by Technical university of Chemnitz University of Technology, Germany. Also, we thank the Helmholtz-Zentrum Berlin für Materialien und Energie for the allocation of synchrotron radiation beamtime. We acknowledge Prof. Emeritus Paul W. Weiblen (University of Minnesota, USA), Royal Ontario Museum (Canada) and Chair of Raw material minerology (Montanuniversitäet Leoben) for providing the mineral specimens of minnesotaite, annite and biotite respectively.

## Conflict of Interest

The authors declare no conflict of interest.

## Author Contributions

M.Z.K. with assistance of A.M. carried out the sample preparation, AFM, and MFM characterization, and with M.A.A. carried out Raman spectroscopy measurements. M.Z.K. with C.T. interpreted AFM and MFM data. O.P. carried out the calculations. M.Z. with support of J.R. performed EPMA/WDS characterization, and WDS data interpretation. A.S. carried out SQUID experiments, and with G.S. interpreted SQUID data. O.S. performed XPS measurements, and with help of G.S. and D.R.T.Z. interpreted the data. M.Z.K. under supervision of S.V. and F.K. performed XAS experiments M.Z.K compiled all the data. M.Z.K., A.M. and O.P. wrote the manuscript. All authors discussed the results and reviewed the manuscript.